\documentstyle{llncs}
\begin{document}
\pagestyle{empty} \mainmatter
\title{Maple+GrTensorII libraries for cosmology}

\titlerunning{Maple+GrTensorII libraries for cosmology}

\author{Dumitru N. Vulcanov\inst{1} \and Valentina D.Vulcanov\inst{2}}

\authorrunning{Dumitru Vulcanov and Valentina Vulcanov}

\tocauthor{Dumitru N. Vulcanov (West University of Timi\c soara)}

\institute{$^1$The West University of Timi\c soara\\
Faculty of Physics, Theoretical and Computational Physics Department\\
\email{vulcan@physics.uvt.ro}\\
$^2$The West University of Timi\c soara\\
Faculty of Mathematics,\\
\email{vali@mitica.physics.uvt.ro}}

\maketitle

\begin{abstract}

The article mainly presents some results in using MAPLE platform 
for computer algebra (CA) and GrTensorII package in doing calculations 
for theoretical and numerical cosmology.


\end{abstract}

\section{Introduction}

Modern cosmology is based on general relativity (GR) and Einstein equations. 
GR requires lengthy (or cumbersome) calculations which could be solved by 
computer algebra methods. During the years, a plethora of CA platforms was 
used for GR purposes, as REDUCE (with EXCALC package), SHEEP or MAXIMA  (see 
for example in \cite{1}, \cite{2} or \cite{3a}). 
Although some advantages as flexibility and 
speed were obvious, recently, platforms as MAPLE or MATHEMATICA are preferred 
by those working in the field, due to their more advanced graphical 
facilities - for a comparison between MAPLE and REDUCE see \cite{3}.

In the last years, an increased interest in theoretical cosmology is visible
because of the new facts revealed by the experimental astrophysics, mainly in
the sense that the universe is actually in an accelerated expansion period -
the so called ``cosmic acceleration'' (see \cite{6}) . In order to fit these 
new facts with
the standard model of the Universe some new mechanisms are proposed, based
on dark-matter, dark-energy and/or cosmological constant hypothesis. New models
are proposed in the literature practically on a daily basis demanding
new specific tools and libraries from the computational science, including
CA applications specially designed for theoretical cosmology. 
Thus we concentrate here 
in symbolic manipulation of Einstein equations with MAPLE and GrTensorII
package (see at {\bf http://grtensor.org}). We packed our procedures in a
specific library, containing all the necessary ingredients for theoretical
cosmology -  Friedmann equations, a scalar field minimally coupled with
gravity and other matter fields terms to be used specifically.

The article is organized as follows :
next section 2 introduces shortly GrTensorII package and his main facilities. 
Then section 3 presents how we implemented non-vacuum Einstein equations in 
a specific form for cosmology (based on Friedmann-Robertson-Walker - 
FRW metric) 
with the stress-energy tensor components designed for interacting with gravity 
matter and one real scalar field separately added. The last section is 
dedicated 
to some new results we obtained with our MAPLE libraries in the so called 
``reverse-technology'' \cite{5} method for treating inflation and cosmic 
expansion triggered by a real scalar field.

Our library, called Cosmo, can be provided by request to the authors. 
We mainly used MAPLE 7 
and MAPLE9 versions but as far as we know the library can be used with other 
MAPLE environments starting with MAPLE V.

\section{Some words about GrTensorII package}

GrTensorII is a free  package from {\bf http://grtensor.org} for the 
calculation 
and manipulation of components of tensors and related objects, embedded in  
MAPLE. Rather than focus upon a specific type or method of calculation, 
the package has been designed to operate efficiently for a wide range of 
applications and allows the use of a number of different mathematical formalisms.  
Algorithms are optimized for the individual formalisms and transformations 
between formalisms has been made simple and intuitive. Additionally, the
package allows for customization and expansion with the ability to
define new objects, user-defined algorithms, and add-on libraries.

The geometrical environment for which GrTensorII is designed is a Riemannian
manifold with connection compatible with the riemannian metric. 
Thus there are special
commands and routines for introducing and calculating geometrical objects as
the metric, Christoffel symbols, curvature (Ricci tensor and scalar)
and the Einstein tensor - as for a couple of examples. Manipulating with 
indices and extracting tensor components are easy to do from some special 
commands and conventions. GrTensorII has a powerful facility for defining new 
tensors, using their natural definitions. As for an example, for calculating 
the Bianchi identities
\begin{equation}\label{bianchi}
G^i_{j;i}=0
\end{equation}
(where $G_{ij}=R_{ij}-\frac{1}{2}g_{ij}R$ is the Einstein tensor defined with
the Ricci tensor $R_{ij}$ and the Ricci scalar $R$, $g_{ij}$ is the metric and
we denoted with the semicolon $;$ the covariant derivative of the metric 
compatible connection)  we can use a short sequence of GrTensorII commands 
for calculating the left side of eq. \ref{bianchi} : 
\begin{verbatim}
> grtw();
> qload(rob_sons);
> grdef(`bia{ ^i }:=G{ ^i ^j ;j }`);
> grcalc(bia(up)); grdisplay(bia(up));
\end{verbatim}
Actually above, the first two commands are for starting the GrTensorII 
package and loading the FRW metric (previously constructed and stored in 
a special directory - GrTensorII provides also an entire collection of 
predefined metrics, but the user can also define his owns using a {\bf 
gmake(...)} command). The last line contains two commands, for effectively 
calculating the new {\bf bia(up)} tensor and for displaying the results. 
If the metric in discussion is compatible with the connection the {\bf bia()}
tensor must have all components vanishing.

The central point of any calculation with GrTensorII is {\bf grcalc()} command.
Often large terms result in individual tensor components which need to be 
simplified. For this {\bf gralter()} and {\bf grmap()} commands are provided 
equiped with several simplifying options, mainly coming from the simplifying 
commands of MAPLE and some specific to GrTensorII. Actually the user is free 
to choose his own simplification strategy inside these commands. 

Special libraries are also available for doing calculation in different frames
or basis and in Newman-Penrose formalism.

\section{The Cosmo library}
As we mentioned earlier, in modern cosmology we are using the 
Friedmann-Robertson-Walker metric (FRW), having the line element in spherical 
coordinates 
\begin{equation}\label{FRW}
ds^2 = -c^2 dt^2 + R(t)^2 \left [ \frac{dr^2}{1-k r^2} + r^2 (d\theta^2 +
\sin^2 \theta~d\phi^2)\right ] 
\end{equation}
as a generic metric for describing the dynamics of the universe. 
Here $k$ is a constant with arbitrary
value, positive (for closed universes), negative (for open universe) and
zero for flat universes. Usually, this constant is taken $1$, $-1$ or
$0$ respectively. $R(t)$ is called scale factor, and is only function of time,
due to the homogeneity and isotropy of space as in standard model of the 
universe is presumed. The dynamic equations are obtained introducing (\ref{FRW}) in the non-vacuum Einstein equations, namely
\begin{equation}\label{EE}
G_{ij}=R_{ij} - \frac{1}{2}g_{ij} R + \lambda g_{ij} = \frac{8 \pi G}
{c^4} T_{ij}
\end{equation}
where $\lambda$ is the cosmological constant, $T_{ij}$ the 
stress-energy tensor, G the gravitational constant, $c$ the speed of light 
and $i,j=0,1,2,3$. The matter content of the universe is given by the
stress-energy tensor $T^{ij}$ which we shall use as :
\begin{equation}\label{tij}
T^{ij} = T^{ij}_{\phi} + T^{ij}_{m}
\end{equation}
where the stress-energy tensor of a scalar field minimally coupled with gravity
and the stress-energy tensor of the matter (other than the scalar field) have
the form of a perfect fluid, namely :
\begin{eqnarray}\label{tphi}
T_{\phi}^{ij} = (p_{\phi}+\rho_{\phi})u^{i}u^{j} + p_{\phi}g^{ij} \\
T_{m}^{ij} = (p +\rho)u^{i}u^{j} + p g^{ij} \label{tmatter}
\end{eqnarray}
Above the scalar field pressure and density are
\begin{eqnarray}\label{pphi}
p_{\phi} = -\frac{1}{2}\partial^{i}\partial_{i} \phi -\frac{1}{2}V(\phi)\\
\label{rhophi}
\rho_{\phi} = -\frac{1}{2}\partial^{i}\partial_{i} \phi +\frac{1}{2}V(\phi)
\end{eqnarray}
Here we used the 4-velocities $u_{i}$ obviously having $u^{i} u_{i}= -1$.

Introducing all these in (\ref{EE}) and defining the Hubble function 
(usually called Hubble constant) and the deceleration factor as
\begin{equation}\label{hubble-acc}
H(t)=\frac{\dot{R}(t)}{R_0}\hbox{~~;~~}Q(t)=-\frac{\ddot{R}(t)}
{2 H(t)^2 R(t)}
\end{equation}
where a dot means time derivative and $R_0$ is the initial (actual) scale
factor, we should obtain the dynamical equations describing the behavior
of the universe, the so called Friedmann equations. The whole package will
contain also the conservation laws equations and the Klein-Gordon equation
for the scalar field, separately. We composed a sequence of GrTensorII
commands for this purpose. First, defining the 4-velocities, the scalar
field functions and the Einstein equations, we have
\begin{verbatim}
> restart;grtw();qload(rob_sons);
> grdef(`u{ i } := -c*kdelta{ i $t}`);
> grdef(`Scal := Phi(t)`);
> grdef(`T1{ i j } := Scal{ ,i }*Scal{ ,j } -
        g{ i j }*(g{ ^a ^b }*Scal{ ,a }*Scal{ ,b }+
                                          V(t))/2`);
> grdef(`TT1{ i j } :=(epsilonphi(t)+
                       pphi(t))*u{ i }*u{ j } +
                       pphi(t)*g{ i j }`);
> pphi(t):=diff(Phi(t),t)^2/2/c^2-V(t)/2;
> epsilonphi(t):=diff(Phi(t),t)^2/2/c^2+V(t)/2;
> grdef(`test{ i j }:=T1{ i j }- TT1{ i j }`);
> grcalc(test(dn,dn)); grdisplay(test(dn,dn)); 
> grdef(`T2{ i j } := (epsilon(t) + p(t))*u{ i }*u{ j } + 
                                          p(t)*g{ i j}`);
> grdef(`T{ i j } :=T1{ i j } + T2{ i j }`);
> grdef(`cons{ i }:= T{ i ^j ;j }`); grcalc(cons(dn));
> EcuKG:=grcomponent(Box[Scal],[]) -DV(t)/2;
> grdef(`Ein{ i j } := G{ i j } - 8*Pi*G*T{ i j }/c^4`);
> grcalc(Ein(dn,dn)); gralter(Ein(dn,dn),expand);
\end{verbatim}
Here we defined twice the stress-energy components for the scalar field,
due to the possibility of a direct definition ({\bf T1()}) and through the
corresponding density and pressure ({\bf TT1()}). Because we are working
in a coordinate frame, these must have equal components and we can check it
through {\bf test(dn,dn)} tensor as having vanishing components. Finally the
total stress-energy tensor and the Einstein equations are defined, 
as it is obvious.
Separately we defined the conservation law-equation ({\bf cons()}) as
the contracted covariant derivative of the stress-energy tensor and the
Klein-Gordon equation for the scalar field - as the unique component of
the d'Alembertian and adding a special function of the derivative of the
potential in terms of the scalar field {\bf DV(t)}. We shall treat this
as an extra variable to be extracted solving the equations. 

Next step is to extract, one by one the components of {\bf Ein(dn,dn)} as
the final form of (\ref{EE}) through a sequence of {\bf grcomponent} commands
followed by certain simplifications and rearrangements of terms. As some
of the equations are identical we shall restrict only to two of them,
coupled with conservation and Klein-Gordon equations. As a result we
denoted with {\bf Ecunr1} and {\bf Ecunr2} the independent Einstein equations
 and with
{\bf Ecunr3} the conservation law equation ({\bf EcuKG} remains as it is).
We also provided a separate equation ({\bf Ecnur22}) for one of the
above terms written with the acceleration factor {\bf Q(t)}. Then comes a
series of substitution commands for casting the equations in terms of 
the Hubble function, deceleration factor and geometrical factor
defined as $K(t)=k/R(t)^2$ :
\begin{verbatim}
> Ecunr1:=expand(simplify(subs(k=K(t)*RR(t)^2,Ecunr1)));
> Ecunr2:=expand(simplify(subs(k=K(t)*RR(t)^2,Ecunr2)));
> Ecunr1:=subs(diff(RR(t),t)=H(t)*RR(t),Ecunr1);
> Ecunr22:=subs(diff(RR(t),t,t)=-2*H(t)^2*RR(t)*Q(t),
                                               Ecunr2);
> Ecunr22:=subs(diff(RR(t),t)=H(t)*RR(t),Ecunr22);
> Ecunr2:=subs(diff(RR(t),t)=H(t)*RR(t),Ecunr2);
> Ecunr2:=expand(Ecunr2);
> Ecunr2:=subs(diff(RR(t),t)=H(t)*RR(t),Ecunr2);
> Ecunr3:=subs(diff(RR(t),t)=H(t)*RR(t),Ecunr3);
> EcuKG:=subs(diff(RR(t),t)=H(t)*RR(t),EcuKG);
\end{verbatim}
Finally we have the Friedmann equations in the form :
\begin{eqnarray}\label{ecuKG}
\frac{1}{c^2}\left [ \ddot{\phi}(t) + 3H(t)\dot{\phi}(t) \right ]+
\frac{1}{2}DV(t)=0
\end{eqnarray}
\begin{eqnarray}
\label{ecunr1}
3 H(t)^2 +3 c^2 K(t)-\frac{4\pi G}{c^4}\left [\dot{\phi}(t)^2 +
c^2 V(t) + 2c^2 \epsilon(t)\right ] =0
\end{eqnarray}
\begin{eqnarray}
2 \dot{H}(t) +3 H(t)^2 +c^2 K(t)+\frac{4\pi G}{c^4}\left [\dot{\phi}(t)^2
~~~~~~~~~~~~~~~~~~~~~\right . \nonumber \\
\label{ecunr2}
\left .~~~~~~~~~~~~~~~~~~~~~~ -c^2 V(t)+2 c^2 p(t)\right ]=0
\end{eqnarray}
\begin{eqnarray}
H(t)^2 (1-4Q(t)) +c^2 K(t) +\frac{4\pi G}{c^4}\left [\dot{\phi}(t)^2 
~~~~~~~~~~~~~~~~~~~\right . \nonumber \\
\label{ecunr22}
\left .~~~~~~~~~~~~~~ -c^2 V(t) + 2 c^2 p(t) \right ]=0
\end{eqnarray}
\begin{eqnarray}
\frac{1}{c^2}\left [\ddot{\phi}(t)\dot{\phi}(t)+3H(t)\dot{\phi}(t)^2\right ]
+\frac{1}{2}\dot{V}(t) +\dot{\epsilon}(t)+~~~~~~~~~~~~~~~\nonumber\\
\label{ecunr3}
~~~~~~~~~~~~~3H(t) (p(t)+\epsilon(t))=0
\end{eqnarray}
These are the classical Friedmann equations (\ref{ecunr1} and \ref{ecunr2},
\ref{ecunr22}) together with Klein-Gordon equation (\ref{ecuKG}) and
the conservation law (\ref{ecunr3}). 
After all these calculations are done we save a MAPLE type library, called
{\bf cosmo.m} through a {\bf save} command. We have to point out here
that there are some new facts around {\bf save} command starting with
MAPLE 6 version, so for this we need to do as :
\begin{verbatim}
> parse(cat("save ",substring(convert([anames(),
                   "cosmo.m"],string),2..-2)),statement);
\end{verbatim}
Having this library stored, every-time one need the above equations, it can
load fast through a {\bf read} command. It provides all the functions
and variables directly without running all the stuff we presented here above.
Thus, the {\bf cosmo.m} library provides all the necessary environment for 
doing calculation within the standard model of cosmology, with FRW metric
and a scalar field and other matter variables included. 
For these last ones there are some functions left undefined ({\bf epsilon(t)}
and {\bf p(t)}) where the user can define
other matter fields than the scalar field to  be included in the 
model - even a second scalar field and/or the cosmological constant as
describing the dark-energy content of the Universe. Thus our library can
be used in more applications than those we presented in the next section.
In the same purpose, we left in the library some of the original equations
unprocessed - having different names - as for example the components of
the Einstein tensor ({\bf Ein(dn,dn)}). Thus the user can finally save his
own library, expanding the class of the possible applications of our 
{\bf cosmo} library.

 As an example, we shall next point out some results we obtained by using 
this library for the so called ``reverse-technology'' \cite{5} 
treatment of inflation triggered by the scalar field.

\section{Some results} 
In the standard treatment of cosmological models with scalar field, it is
prescribed a certain potential function for the scalar field (taking
into account some physical reasons specific to the model processed) and
then the dynamical Friedmann equations are solved (if it is possible) to
obtain the time behavior of the scale factor of the universe. As recently
some authors pointed out, a somehow ``reverse'' method \cite{5}
is also interesting,
where the time behavior of the scale factor is ``a priori'' prescribed
(as a function of time which will model the supposed time behavior
of the universe in inflation or in cosmic accelerated expansion) then
solving the Friedmann equations we can extract the shape of the corresponding
potential for the theory. This is the so called ``reverse technology'' and
we shall use it here to illustrate the usage of our {\bf cosmo.m} library.

We shall concentrate ourselves to the case of no matter variables other
than the scalar field. In this case we solve first equations (\ref{ecunr1})
and (\ref{ecunr2}) for the potential $V(t)$ and $\dot{\phi}(t)^2$, not
before denoting the last one with a special intermediate Maple function
called {\bf D2Phi(t)} with {\bf subs} command :
\begin{verbatim}
> Ecunr1:=subs(diff(Phi(t),t)^2=D2Phi(t),Ecunr1);
> Ecunr2:=subs(diff(Phi(t),t)^2=D2Phi(t),Ecunr2);
> solve({Ecunr1,Ecunr2},{V(t),D2Phi(t)});
\end{verbatim}
Thus we have \cite{4} :
\begin{equation}\label{pot}
V(t)= \frac{1}{4\pi}\left [ \dot{H}(t) + 3H(t)^2 + 2K(t)\right ]
\end{equation}
\begin{equation}
\label{dotphi2}
\dot{\phi}^2 = \frac{1}{4\pi}\left [-\dot{H}(t) +K(t)\right ]
\end{equation}
Here and in the following pages we have, as usual geometrical units 
$G=c=1$. 
Here we shall process one of the examples pointed out in Ellis and Madsen
article \cite{4}, namely that one of  de Sitter exponential expansion, where
\begin{equation}
R(t) = R_0 e^{\omega t}\hbox{~~~;~~~} H(t) = \omega
\end{equation}
Thus (\ref{pot}) and (\ref{dotphi2}) became
\begin{equation}
V(t)= \frac{3 \omega^2}{4 \pi}+\frac{k}{2\pi e^{2\omega t}} \hbox{~~~;~~~}
\dot{\phi}(t)^2 = \frac{k}{4\pi e^{2\omega t}}
\end{equation}
after simple evaluations of the corresponding Maple expressions. It is obvious
that $\dot{\phi}(t)$ can be simply obtained by square root of the above
expression and can also be integrated to give the potential as:
\begin{equation}\label{phi}
\phi(t) = -\frac{1}{2}\frac{\sqrt{k} e^{-\omega t}}{\sqrt{\pi}\omega}+\phi_0
\end{equation}
The result is that, after evaluating the Einstein equations we have
automatically satisfied {\bf Ecunr1}, {\bf Ecunr2} and {\bf Ecunr3} and
the Klein Gordon equation has the form :
\begin{equation}
{\bf EcuKG} = \frac{\sqrt{k}\omega}{\sqrt{\pi}e^{\omega t}}+\frac{1}{2}
DV(t)=0
\end{equation}
The last one is used to express the {\bf DV(t)} by solving it, and it is
a point to check the calculation if this expression is equal to that one
obtained directly from the potential. But this checking can be done only
if we express, after a sequence of simple {\bf subs} and {\bf solve}
commands, the potential $V(t)$ and his derivative $DV(t)$ in terms of the 
scalar field, more precisely in terms of $\phi(t) - \phi_0$. The result is
\begin{equation}
V(\phi(t)) = \frac{3\omega}{4\pi}+2\omega^2 (\phi(t)-\phi_0)^2
\end{equation}
\begin{equation}
DV(\phi(t))=4\omega^2 (\phi(t)-\phi_0)
\end{equation}
These results are in perfect agreement with the well-known results from
\cite{4}. 

We processed in the same way more examples, some of them
completely new. Our purpose was to produce Maple programs for processing the
``reverse-technology'' \cite{4}-\cite{5}
method for these type of potentials with matter
added to the model, especially dust or radiative matter. Although the steps
for computing are the same, there are two points of the calculations where
troubles can appear and the solution is not straightforward.
The first one is the integration of the {\bf DPhi(t)}
obtained as the square root of {\bf D2Phi(t)}. Sometime it is not trivial
to do this, so in several cases we used approximation techniques, by 
evaluating the cosmological functions at the initial time. Our main purpose
was to produce good initial data for numerical solving the Einstein
equations (with the Cactus code, for example) thus these approximations can
be a good solution for short time after the initial time. The second trouble
point is to evaluate the potential in terms of the scalar field, namely
to extract the time variable from it. Sometimes here we have  
transcendental equations and again some approximation methods can solve the
problem. Because these results are not in the topic of this article
we plan to report them in a future article.

\section*{Acknowledgments}
Special thanks to one of the referees who revealed many week points of 
our article. This work was partially supported by the Romanian Space
Agency (grant nr. 258/2002-2004) and the Albert Einstein Institute,
Potsdam, Germany.

\end{document}